# Field-induced ferromagnetic phase transition in 2D Fermi systems with magnetic dipole-dipole interaction

G. H. Bordbar[1,2], F. Pouresmaeeli[1] and A. Poostforush[1]

[1] *Physics Department, Shiraz University, Shiraz 71454, Iran*[1]

and

[2] *Department of Physics and Astronomy, University of Waterloo, 200 University Avenue West, Waterloo, Ontario, N2L 3G1, Canada*

**Abstract**

Magnetic properties of the two-dimensional spin-polarized Fermi gas with dipole-dipole interaction are studied in the presence of external magnetic field at zero temperature. Within perturbation theory and the second quantization formalism, the total energy is explicitly obtained as a function of three dimensionless parameters, the spin polarization, dipolar coupling and Zeeman parameters. We examine the effects of these agents on the magnetic properties of 2D Fermi gas. The results show that an induced ferromagnetic phase transition is observed only for adequately large values of magnetic field. This paper offers two controllable factors to change the spin polarization of the system.

Keywords: dipole-dipole interaction, two-dimensional Fermi gas, second quantization formalism, magnetic properties, perturbation theory.

**Introduction**

The Fermi gas as one of the fundamental models in many-body physics, has been investigated in condensed matter for past decades [1-3]. In this model, the long-range interactions between fermions determine various observed physical phenomena [4, 5]. Due to the realization of the Bose-Einstein condensate (BEC) for $^{53}Cr$ atoms [6], the dipole-dipole interaction (DDI) has attracted considerable attention. Unlike the Coulomb interaction, dipole-dipole interaction is a non-central interaction which can be controlled both its strength and sign [7].

In Fermi systems, magnetic dipole moments (or spin of particles) interact via dipole-dipole interaction with anisotropic and long-range characteristics which decay faster than the Coulomb interaction at large distance. Due to its particular nature, the DDI has opened new perspectives to examine the many-body systems [8, 9]. It is mentioned that the DDI appears in a strong interaction with highly magnetic Fermionic atoms such as $^{53}Cr$ [6, 10], $^{167}Er$ [11], and $^{161}Dy$ or $^{163}Dy$ [12]. Various associated effects have been discussed in the theoretical literatures. These include stability and excitations of dipolar gases [13, 14], superfluidity in bilayer and multilayer systems [15-17] and topological superfluidity in 2D systems [18, 19].

---

[1] Permanent address

In recent years, numerous theoretical studies have been done to clarify the characteristics of the DDI. The magnetic properties of nucleonic systems with tensor force which has a similar structure to the DDI were determined on the basis of Landau-Fermi liquid theory [20, 21]. In addition, its impact on Fermi systems with spin degrees of freedom was examined via variant approaches. The ground state energy of three-dimensional (3D) Fermi gas with spin-$1/2$ dipolar atoms was found as a function of the average inter-particle distance by Mahanti and Jha [22-24] employing Hartree-Fock approach by considering the spheroid occupation function. Subsequently, theoretical studies on the 3D Fermi gas with dipolar and short-range interactions were carried out within the Hartree-Fock approximation. Especially, Fregoso *et al.* [25, 26] showed that biaxial nematic and ferronematic phases occurred in Fermi gases of dipolar atoms. Based on the perturbation theory, the magnetic properties of Fermi gas with dipole force were also achieved theoretically at zero temperature [27]. On the other hand, the stability of unpolarized 3D dipolar Fermi gas was studied through mean-field theory [28]. In our earlier paper [29], the DDI energy of 3D electron gas was obtained by microscopic analysis. In that work, it was shown that the energy of this system was presented by the summation of all energy contributions of the states with opposite parity.

Several efforts have been made to study the ground state properties of 2D dipolar Fermi gas. At weak coupling, the Fermi-liquid properties of 2D dipolar Fermi gas was computed applying the perturbation theory by Lu and Shlyapnikov [30]. These authors expressed thermodynamic quantities as a power series up to second order in the dimensionless dipolar parameter. In another theoretical study, Matveeva and Giorgini [31] employed the Quantum Monte Carlo method to obtain the numerical results for the phase diagram of dipolar Fermi gas over a wide range of dipolar parameter. Additionally, within the Euler-Lagrange Fermi-hypernetted-chain approximation, theoretical studies on the ground-state properties of dipolar Fermi fluid were performed by Abedinpour and coworkers [32].

Since the polarization rate of system is very sensitive to the strength of magnetic field, controlling the magnitude of external field becomes consequential. Moreover, sudden changes in the response function to the magnetic field lead to the emergence of phase transition. Therefore, the magnetic field acts as an effective factor for creating the ferromagnetic order and tuning the spin polarization of the system. Despite numerous studies on the effect of DDI in Fermi systems, including dipolar Fermi gas, there have been no reviews yet to investigate the effect of external magnetic field on these polarized systems. In the present work, we investigate the magnetic properties of a two-dimensional polarized Fermi gas with spin $1/2$ at zero temperature. The chargeless Fermionic particles interacting via the dipole-dipole interaction are subjected to an external uniform magnetic field along the *z* axis. Employing the perturbation theory, the analytic expression for the ground state energy at weakly coupling regime is obtained. The numerical results are in reasonable agreement with theoretical calculations. In this approach based on the second quantization formalism, the DDI energy is represented as a sum of partial energies with even and odd quantum numbers. Finally, the effect of magnetic field and dipolar parameter on the magnetic properties of system are reported.

**Model and Theory**

We consider a uniform 2D Fermi gas consisting of $N$ Fermionic particles with mass $m$, spin $1/2$ and magnetic dipole moment $\mathbf{d} = d_0 \mathbf{S}$, where $\mathbf{S}$ is the spin operator and $d_0$ is the strength of magnetic dipole moment. By applying the uniform magnetic field along the z direction, the spin-polarized system subsequently consists of $n_+$ parallel and $n_-$ antiparallel spins with respect to the magnetic field direction where $n_\sigma$ denotes the number density for spin $\sigma = +, -$.

The spin polarization as the most used parameter is defined as

$$\xi = \frac{n_+ - n_-}{n}, \tag{1}$$

where $-1 \leq \xi \leq 1$, and $n = n_+ + n_-$ is the total number density of the system.

For this system, the total Hamiltonian is described as

$$\hat{H} = -\frac{\hbar^2}{2m} \sum_{i=1}^{N} \nabla_{\mathbf{r}_i}^2 + \sum_{i=1}^{N} \mathbf{d}_i \cdot \mathbf{B} + \hat{H}_{d-d}, \tag{2}$$

where the first and the second terms are the kinetic energy and Zeeman energy of the Fermi particles.

The magnetic dipole-dipole interaction (DDI) between two particles with dipole moments $\mathbf{d}_i$ and $\mathbf{d}_j$ is as follows,

$$\hat{H}_{d-d} = \frac{1}{2} \frac{\mu_0}{4\pi} \sum_{i \neq j=1}^{N} \frac{\mathbf{d}_i \cdot \mathbf{d}_j - 3(\mathbf{d}_i \cdot \hat{\mathbf{r}})(\mathbf{d}_j \cdot \hat{\mathbf{r}})}{|\mathbf{r}_i - \mathbf{r}_j|^3}. \tag{3}$$

In the above equation, $\mathbf{r} = \mathbf{r}_i - \mathbf{r}_j$ and $\hat{\mathbf{r}} = \mathbf{r}/|\mathbf{r}|$ are relative position of particles and its unit vector, respectively. For this interacting many-body system, the dipolar interaction is presented in the second quantization formalism [33] for any quantum state (denoted by $\beta \equiv \mathbf{k}, \sigma$) as follows,

$$\hat{H}_{dd} = \frac{1}{2} \sum_{\beta_1 \beta_2 \beta_3 \beta_4} \langle \beta_1, \beta_2 | \hat{V}_{dd} | \beta_3, \beta_4 \rangle a^\dagger_{\beta_1} a^\dagger_{\beta_2} a_{\beta_4} a_{\beta_3}. \tag{4}$$

In the above equation, the two-body dipolar interaction is given by

$$\hat{V}_{dd} = \frac{C_{dd}}{4\pi} \frac{\hat{S}_{12}}{\mathbf{r}^3}, \tag{5}$$

where

$$C_{dd} = \mu_0 d_0^2, \qquad \hat{S}_{12} = (\mathbf{S}_1 \cdot \mathbf{S}_2) - 3(\mathbf{S}_1 \cdot \hat{\mathbf{r}})(\mathbf{S}_2 \cdot \hat{\mathbf{r}}). \tag{6}$$

For a 2D Fermi gas, one can calculate the matrix elements of Eq. (4) with respect to the single-particle wave function expressed by a plane wave function and spin state with z-component, $\chi_\sigma$, in the surface area of $A$,

$$\psi_{\mathbf{k},\sigma}(\mathbf{r}) = \langle \mathbf{r} | \mathbf{k},\sigma \rangle = A^{-\frac{1}{2}} e^{i\mathbf{k}\cdot\mathbf{r}} \chi_\sigma \tag{7}$$

Using the following variables, the two-body system can be expressed in the center-of- mass and relative coordinates,

$$\mathbf{K} = \mathbf{k}_1 + \mathbf{k}_2, \qquad \mathbf{K}' = \mathbf{k}_3 + \mathbf{k}_4, \qquad \boldsymbol{\kappa} = \frac{\mathbf{k}_1 - \mathbf{k}_2}{2}, \qquad \boldsymbol{\kappa}' = \frac{\mathbf{k}_3 - \mathbf{k}_4}{2}. \tag{8}$$

Regarding the singularity of dipolar interaction at the origin, we use the expansion of plane wave in terms of cylindrical waves to simplify the calculations.

$$e^{i\mathbf{k}\cdot\mathbf{r}} = \sum_{m=-\infty}^{+\infty} i^{-m} J_m(k_\rho \rho) e^{-im\phi_k} e^{im\phi}. \tag{9}$$

In Eq. (4), the two-body matrix elements are obtained as

$$V(\boldsymbol{\kappa},\boldsymbol{\kappa}') = \langle \mathbf{k}_1 \sigma_1, \mathbf{k}_2 \sigma_2 | \hat{V}_{dd} | \mathbf{k}_3 \sigma_3, \mathbf{k}_4 \sigma_4 \rangle$$
$$= \frac{2\pi}{A} \frac{C_{dd}}{4\pi} \sum_{\substack{SM_s \\ S'M_s'}} F(\boldsymbol{\kappa},\boldsymbol{\kappa}') V_{m,SM_s;m',S'M_s'} \{C_{SM_s}^{\sigma_1\sigma_2} C_{S'M_s'}^{\sigma_3\sigma_4}\} \delta_{\mathbf{KK}'}, \tag{10}$$

where

$$F(\boldsymbol{\kappa},\boldsymbol{\kappa}') = \sum_{m,m'} e^{im\phi_\kappa} e^{-im'\phi_{\kappa'}} \int i^{m-m'} J_m(\kappa\rho) \frac{1}{\rho^3} J_{m'}(\kappa'\rho) \rho d\rho. \tag{11}$$

In this expression, $J_m(\kappa\rho)$ is the Bessel function of order $m$ with wave number $\kappa$ and $C_{SM_s}^{\sigma_1\sigma_2}$ is the Clebsch-Gordan coefficient corresponding to the addition of $\sigma_1$ and $\sigma_2$. Also $V_{m,SM_s;m',S'M_s'} = \langle m, SM_s | \hat{S}_{12} | m', S'M_s' \rangle$ are the matrix elements of $\hat{S}_{12}$.

Accordingly, the two-body dipolar Hamiltonian in the second quantization form is found as

$$\hat{H}_{dd} = \frac{1}{2} \sum_{\mathbf{kpq}} \sum_{\substack{\sigma_1\sigma_2 \\ \sigma_3\sigma_4}} V(\boldsymbol{\kappa},\boldsymbol{\kappa}') a_{\mathbf{k+q},\sigma_1}^\dagger a_{\mathbf{p-q},\sigma_2}^\dagger a_{\mathbf{p},\sigma_4} a_{\mathbf{k},\sigma_3}, \tag{12}$$

where

$$\boldsymbol{\kappa}' = \frac{\mathbf{k} - \mathbf{p}}{2}, \qquad \boldsymbol{\kappa} = \boldsymbol{\kappa}' + \mathbf{q}. \tag{13}$$

The ground state properties of dipolar Fermi gas are characterized by three parameters, spin polarization, $\xi$, the dimensionless dipolar coupling parameter, $\lambda$, and the dimensionless Zeeman parameter, $\Lambda$. The dipolar and Zeeman parameters can be expressed as

$$\lambda = \frac{C_{dd}}{8\pi \varepsilon_f} k_f^3, \qquad \Lambda = \frac{2d_0 B}{\varepsilon_f} \qquad (14)$$

where $\varepsilon_f$ and $k_f = \sqrt{4\pi n}$ are the Fermi energy and Fermi wave number of full-polarized 2D Fermi gas, respectively.

For small dipolar parameter, the kinetic energy has a dominant effect compared to dipole-dipole interaction. In other word, the DDI interaction behaves as a perturbative effect at weak coupling. Under the aforementioned condition, the perturbation theory can be applied as an appropriate method to calculate the first-order energy. Consequently, one can compute the DDI energy by computing the expectation value of Hamiltonian in the normalized ground state, $|F\rangle$,

$$E^{(1)} = \langle F | \hat{H}_{d-d} | F \rangle \qquad (15)$$

The normalized ground state is only occupied by states with momenta less than the Fermi wave number. Therefore, the following conditions should be taken into account,

$$\mathbf{k} + \mathbf{q} = \mathbf{p}, \quad \sigma_1 = \sigma_4, \quad \sigma_2 = \sigma_3. \qquad (16)$$

This makes it possible to reach the non-zero value of the dipolar energy. It is also worth mentioning that the direct term ($q = 0$) has no contribution in the two-body energy.

At the zero temperature, the distribution function of particles with spin $\sigma$ corresponding to a circular Fermi surface is given by

$$n_{\sigma,\mathbf{k}} = \langle a^\dagger_{\sigma\mathbf{k}} a_{\sigma\mathbf{k}} \rangle = \Theta(k_{f\sigma} - k) \qquad (17)$$

where $\Theta(x)$ is the step function.

The Fermi wave number can be determined through the expectation value of the number operator,

$$n_\sigma = A^{-1} \sum_{\mathbf{k}} \langle a^\dagger_{\sigma\mathbf{k}} a_{\sigma\mathbf{k}} \rangle$$

$$= A^{-1} \sum_{\mathbf{k}} \Theta(k_{f\sigma} - |\mathbf{k}|) = \frac{k_{f\sigma}^2}{4\pi} \qquad (18)$$

The resulting expression for the DDI energy is as

$$E^{(1)} = -\frac{C_{dd}}{4\pi A} \sum_{\mathbf{kq}} \sum_{\sigma_1 \sigma_2} \sum_{m=-\infty}^{+\infty} \sum_{SM_s} \frac{2q(-1)^{1-S+m}}{4m^2-1} \left| C^{\sigma_1 \sigma_2}_{SM_s} \right|^2 V_{SM_s;SM_s} \qquad (19)$$
$$\Theta(k_{f\sigma_1} - |\mathbf{k}+\mathbf{q}|)\Theta(k_{f\sigma_2} - |\mathbf{k}|)$$

Substituting Eq. (18) into Eq. (19), and employing the diagonal matrix elements of $\hat{S}_{12}$ and the Clebsch-Gordan coefficients, the first-order perturbation energy of DDI can be written as

$$E^{(1)} = \frac{C_{dd}}{4\pi} \frac{A}{16\pi} \left[ \frac{128}{45\pi} (k_{f+}^5 + k_{f-}^5) - 2k_f^5 \{I(\xi) + h(\xi)\} \right], \qquad (20)$$

Where

$$I(\xi) \equiv \frac{1}{\pi k_f^5} \sum_{\sigma=+,-} \int_{|k_{f+}-k_{f-}|}^{k_{f+}+k_{f-}} 2q^2 (k_{f\sigma}^2 \operatorname{Arcsec}(\frac{k_{f\sigma}}{q_\sigma}) - q_\sigma \sqrt{k_{f\sigma}^2 - q_\sigma^2}) dq ,\tag{21}$$

$$h(\xi) \equiv \frac{2}{3k_f^5} \begin{cases} k_{f-}^2(k_{f+}-k_{f-})^3 & k_{f+} > k_{f-} \\ k_{f+}^2(k_{f-}-k_{f+})^3 & k_{f+} < k_{f-} \end{cases},\tag{22}$$

$$q_\pm = \frac{q^2 \pm k_{f+}^2 \mp k_{f-}^2}{2q},\tag{23}$$

and

$$k_{f\pm}^2 = 2\pi n (1\pm \xi).\tag{24}$$

The contributions of kinetic energy and Zeeman energy per particle are also given as

$$\frac{E^{(0)}}{N} = \frac{\hbar^2 \pi n}{2m}(1+\xi^2), \qquad \frac{E_M}{N} = -d_o B \xi .\tag{25}$$

Therefore, the total energy per particle of 2D Fermi gas with the dipole-dipole interaction in terms of the dimensionless parameters can be expressed as

$$\begin{aligned}\frac{E}{N} &= \frac{E^{(0)}}{N} + \frac{E_M}{N} + \frac{E^{(1)}}{N} \\ &= \varepsilon_0 \left[ \frac{1}{2}(1+\xi^2) - \Lambda \xi + \lambda \left\{ \frac{128}{45\pi} \{ (\frac{1+\xi}{2})^{\frac{5}{2}} + (\frac{1-\xi}{2})^{\frac{5}{2}} \} - 2\{I(\xi) + h(\xi)\} \right\} \right]\end{aligned}\tag{26}$$

where $\varepsilon_0 = \frac{\varepsilon_f}{2} = \frac{\hbar^2 k_f^2}{4m} = \frac{\hbar^2 \pi n}{m}$.

In Eq. (26), the second term in the bracket reaches zero value when the 2D Fermi gas is fully polarized. Therefore, the circular Fermi surface is a suitable approximation for this case. It should be noted that the deformation of Fermi surface stems from the anisotropic nature of dipole-dipole interaction. Consequently, it is expected that by considering the deformation in the distribution function, the value of DDI energy slightly is changed.

**Results and Discussion**

The total energy per particle (Eq. (26)) is calculated to compute the magnetic properties of spin-polarized 2D Fermi gas with the dipole-dipole interaction subjected to an external magnetic field. In this section, the numerical results are presented.

In Fig. 1, the total energy per particle of 2D Fermi gas versus spin polarization parameter for various values of $\Lambda$ at $\lambda = 0.5$ is plotted. It can be seen that the energy is reduced with increasing the Zeeman parameter, and the 2D Fermi gas becomes more stable. It is obvious that the energy slightly changes for the Zeeman parameters with values less than about 0.5, and consequently the corresponding magnetic field does not get a substantial influence. Furthermore, the symmetry of total energy with respect to spin polarization is broken in the presence of magnetic field, and the minimum value of the energy of system occurs at the non-zero value of spin polarization parameter in $0 < \xi \leq 1$. At $\lambda = 0.5$, we have found that for the Zemman parameter $\Lambda \lesssim 2.0$, the value of spin polarization corresponding to the minimum point of the energy is less than +1, and so the system is partially polarized. However, the minimum of energy (ground state energy) happens at the polarization of +1 for larger values of Zeeman parameter, and the system reaches the saturation point. This means that the ferromagnetic state is available, and 2D Fermi gas is fully polarized. It should be noted that this threshold limit of Zeeman parameter increases as the dipolar parameter enhances.

The partial DDI energy per particle of spin-polarized Fermi gas corresponding to the azimuthal quantum number, $m$, at $\lambda = 0.5$ and $\Lambda = 1$ is listed in Table 1. The DDI energy is found to be a negative (positive) value for the even (odd) values of $m$, except in the case of $m = 0$.

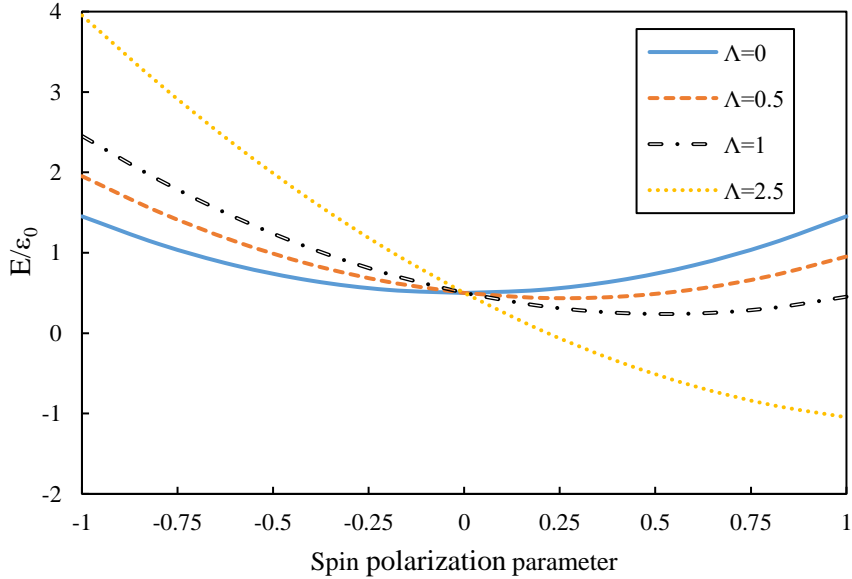

Fig. 1. The total energy per particle (in units of $\varepsilon_0$) versus spin polarization parameter for various values of $\Lambda$ at $\lambda = 0.5$.

Invariance of the DDI against the spatial inversion leads to the conservation of the parity, where the spatial wave functions with even (odd) azimuthal quantum number have even (odd) parity. The factor $(-1)^m$ in Eq. (19) indicating the parity of system, confirms this result. This behavior has been also reported for tensor force in two-nucleon systems. As can be expected, with an increase in values of $m$, the magnitude of DDI energy decreases, and it approaches a specific value.

Table 1. The partial DDI energy per particle (in units of $\varepsilon_0$) corresponding to the azimuthal quantum number $m$, at $\lambda = 0.5$ and $\Lambda = 1$.

| Azimuthal quantum number (m) | Partial DDI energy per particle |
|---|---|
| 0 | 0.079391 |
| ±1 | 0.026464 |
| ±2 | −0.00529 |
| ±3 | 0.002268 |
| ±4 | −0.00126 |

The ground state energy per particle of polarized 2D Fermi gas versus dipolar parameter is shown for some values of Zeeman parameter in Fig. 2. It is seen that with enhancement of dipolar parameter, the energy per particle for each Zeeman parameter increases. By increasing the Zeeman parameter, the rate of change of energy increases at each dipolar parameter. This indicates that an increase of the magnitude of magnetic field induces growth of the increasing rate of Zeeman energy.

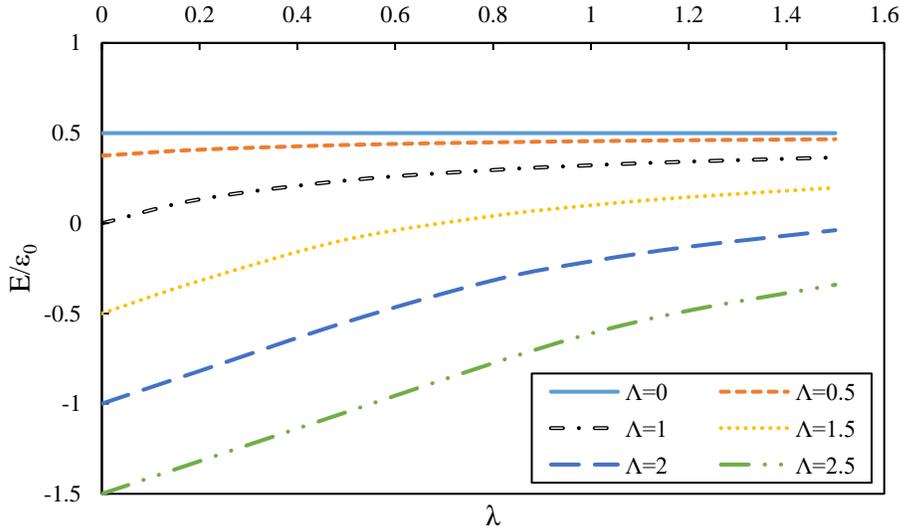

Fig. 2. The ground state energy per particle of polarized 2D Fermi gas (in units of $\varepsilon_0$) versus dipolar parameter for some values of Zeeman parameter.

In Fig. 3, the variation of ground state energy per particle with respect to Zeeman parameter for different values of dipolar parameter is plotted. With an increase of Zeeman parameter up to $0.5$, the energy nearly remains constant. However, the energy is dropped drastically for values of $\Lambda \gtrsim 0.5$. This behavior reveals that the magnetic field is an efficient factor in this range. It is also seen that with the growth of dipolar parameter, the magnitude of energy per particle increases for a distinct Zeeman parameter.

For the equilibrium state, the spin polarization parameter (or dimensionless magnetization) as a function of dipolar parameter for various Zeeman parameter is presented in Fig. 4. It is observed that for each value of $\Lambda$, the spin polarization increases with decreasing the dipolar parameter. This point stems from the fact that as dipolar parameter decreases, the inter-particle distance of fermions increases leading to a reduction in repulsion between the particles. Therefore, it is not necessary to adhere to the Pauli exclusion principle, and spins tend to orient in parallel. In other word, the probability of existence of ferromagnetic state is more than the paramagnetic state. For small value of Zeeman parameter, the Fermi gas is nearly unpolarized. Conversely, at large value of this parameter, the system has a high polarization even at higher values of $\lambda$. Consequently, the magnetic field plays a substantial role against the dipolar interaction in polarization of this system.

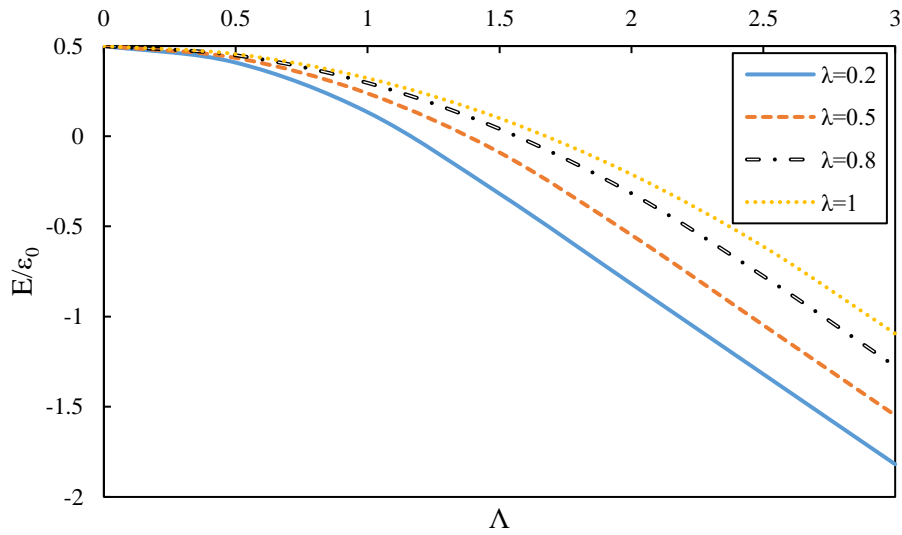

Fig. 3. The variation of the ground state energy per particle (in units of $\varepsilon_0$) versus Zeeman parameter for different values of dipolar parameter.

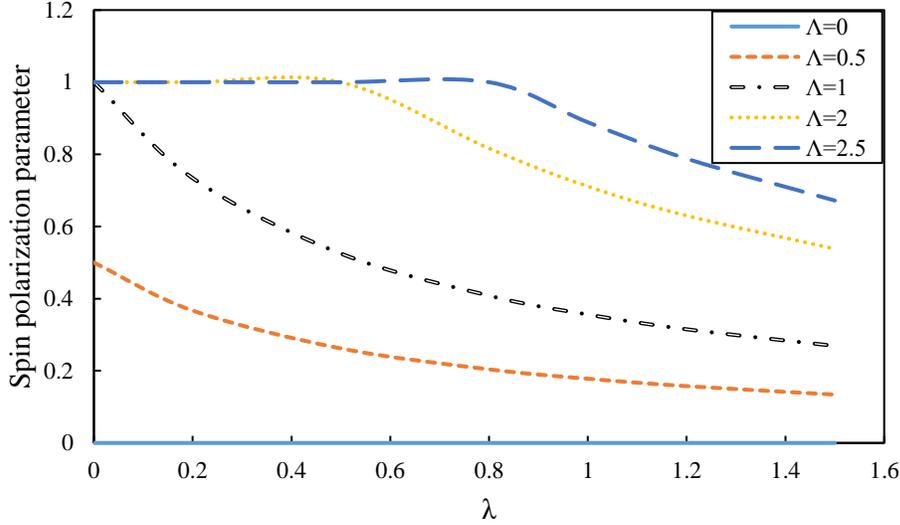

Fig. 4. The spin polarization parameter corresponding to the equilibrium state as a function of dipolar parameter for various Zeeman parameter.

The variations of kinetic energy, the DDI energy and Zeeman energy per particle are separately shown with respect to dipolar parameter at $\Lambda = 2.5$ in Fig. 5. In perturbative regime, the contribution of kinetic energy is relatively large in comparison to the contribution of DDI energy. Furthermore, the Zeeman energy substantially has a dominant effect compared to others. The numerical results at $\lambda \leq 0.8$ which represent a fully polarized Fermi gas in Fig. 4 are in a reasonable agreement with the results of full-polarized 2D dipolar gas reported in Refs. [30-32].

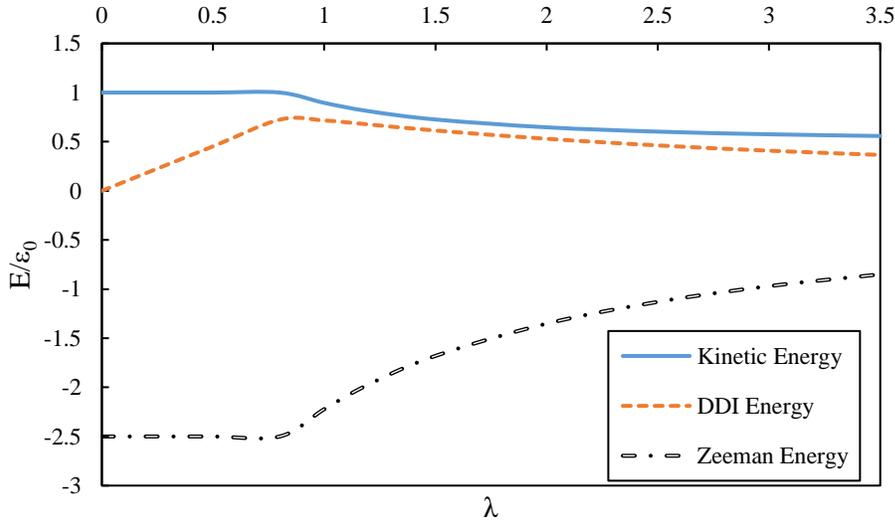

Fig. 5. The variation of kinetic energy, the DDI energy and Zeeman energy (in units of $\varepsilon_0$) versus dipolar parameter at $\Lambda = 2.5$.

In Fig. 6, the influence of Zeeman parameter on the spin polarization parameter (dimensionless magnetization) at equilibrium state for various values of $\lambda$ is illustrated. For small value of $\Lambda$, the spin polarization parameter almost reaches zero where it demonstrates spin symmetry of 2D Fermi gas. With the growth of Zeeman parameter, 2D Fermi gas can be partially polarized, and the spin polarization of the system is maximized for small values of $\lambda$. Moreover, the value of spin polarization parameter increases rapidly with an increase in Zeeman parameter. This phenomenon reveals the existence of an induced ferromagnetic phase transition in the presence of a strong magnetic field. It should be noted that for any value of dipolar parameter, the value of spin polarization corresponding to the large value of $\Lambda$ is equal to +1, and the ferromagnetic state becomes more expectable. Accordingly, the magnetic field acts as a symmetry breaking that generates the ferromagnetic order. In addition, the threshold limit of Zeeman parameter enhances with increasing the dipolar parameter.

The response of a system to the magnetic field, the magnetic susceptibility, is defined as

$$\chi(n,B) = (\frac{\partial M(n,B)}{\partial B})_n$$

Fig. 7 depicts the dimensionless magnetic susceptibility, $\chi / \frac{Nd_0^2}{\varepsilon_0}$, as a function of Zeeman parameter at four values of dipolar parameter. It is clear that for each $\lambda$, a maximum point occurs at the specific Zeeman parameter ($\Lambda_m$) in this curve. This result is found to be a direct evidence for a ferromagnetic phase transition induced by external magnetic field. Moreover, the maximum point of Zeeman parameter corresponding to the phase transition point clearly depends on the strength of the DDI.

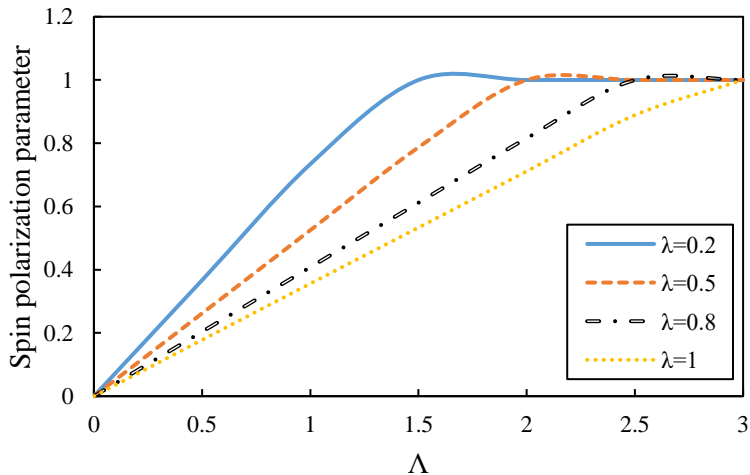

Fig. 6. The influence of Zeeman parameter on the spin polarization parameter at the equilibrium state for various values of $\lambda$.

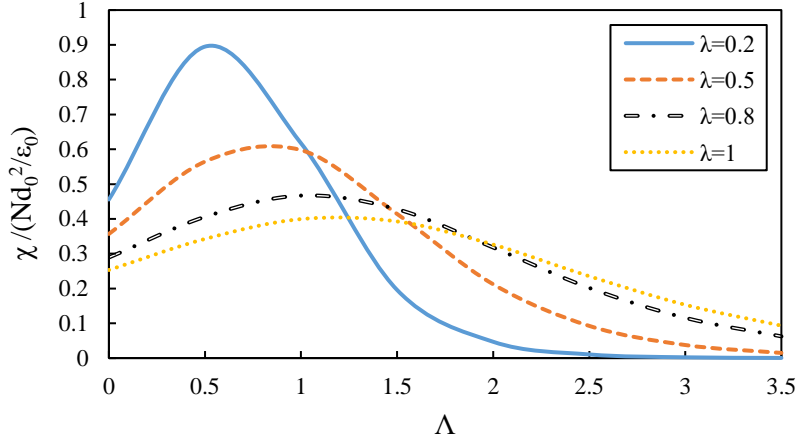

Fig. 7. The dimensionless magnetic susceptibility ($\chi / \frac{Nd_0^2}{\varepsilon_0}$) as a function of Zeeman parameter at four values of dipolar parameter.

In Fig. 8, the phase diagram shows the ferromagnetic and paramagnetic phases separated by a single line. It can be observed that $\Lambda_m$ enhances monotonically with an increase in dipolar parameter. It is concluded that in stronger dipolar coupling, the induced phase transition is observed at larger values of magnetic field. To sum up, the results show that the spin polarization of the system can be changed through two agents, the magnetic field and the dipolar coupling parameter. The value of the dipolar coupling parameter can be controlled by number density and particular choices of atoms.

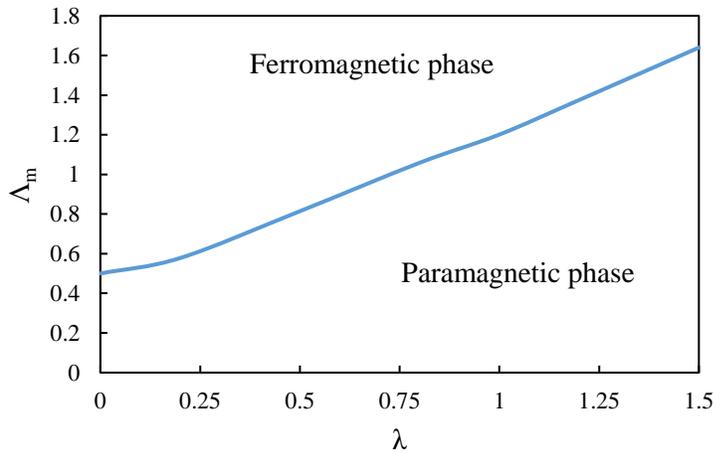

Fig. 8. Phase diagram for the spin-polarized 2D Fermi gas with dipole-dipole interaction in the presence of magnetic field.

## Summary and Conclusion

In summary, the magnetic characteristics of 2D polarized Fermi gas with the dipole-dipole interaction subjected to an external magnetic field were investigated employing the perturbation theory. In the framework of second quantization formalism, the total energy was calculated in terms of spin polarization, dipolar and Zeeman parameters. The results of our investigation are quite convincing in comparison with those reported for 2D fully polarized dipolar Fermi gas [30-32]. The DDI energy was also represented as a sum of partial energy of states with opposite parities. The results indicated that for $\Lambda < 0.5$, the magnetic field has a nearly negligible effect. When the magnetic field is applied, the minimum energy becomes available at $0 < \xi \leq 1$ depending on the strength of the magnetic field. As a consequence, the magnetic field causes the symmetry breaking, and creates the ferromagnetic order. By increasing Zeeman parameter, the ground state energy decreases, and gives rise to a more stable system. It is evident that the effect of the magnetic field on the magnetic properties of 2D Fermi gas becomes more significant and visible when the Zeeman parameter is greater than $0.5$. By increasing the Zeeman parameter, the rate of changes of energy increases for any dipolar parameter. For a fixed Zeeman parameter, it was found that the ground state energy grows with increasing the dipolar parameter. Furthermore, the spin polarization increases as the dipolar parameter decreases. Therefore, we can conclude that the dipole-dipole interaction plays a weak role on the magnitude of spin polarization of this system with respect to that of the magnetic field. It was seen that an induced ferromagnetic phase transition occurs in the presence of magnetic field, and for larger dipolar parameters, this phenomenon becomes observable in the stronger magnetic field. Finally, based on the results of this research, the dipolar coupling parameter and the magnetic field were suggested as controllable means of changing spin polarization for 2D Fermi gas.

## Acknowledgements

We wish to thank Shiraz University Research Council. G. H. Bordbar also wishes to thank Physics Department of University of Waterloo for the great hospitality during his sabbatical.